\documentclass[a4paper, 10pt, twocolumn]{article}

\usepackage[utf8]{inputenc}         
\usepackage[english]{babel}          

\usepackage{graphicx}               
\usepackage{tabularx}               
\usepackage{multirow}               
\PassOptionsToPackage{hyphens}{url}\usepackage{url}                

\usepackage[small,bf]{caption2}     
\usepackage{parskip}
\usepackage{titlesec}
\usepackage{amsmath}    
\usepackage{hyperref}
\usepackage{booktabs}
\usepackage{pdflscape}
\usepackage[sort,compress]{cite}

\titleformat{\section}{\normalfont\large\bfseries}{\thesection}{}{}
\titleformat{\subsection}{\normalfont\large\bfseries}{\thesection}{}{}
\titleformat{\paragraph}{\normalfont\bfseries}{\theparagraph}{}{}
\titlespacing{\section}{0pt}{6pt}{-1pt}
\titlespacing{\subsection}{0pt}{3pt}{-1pt}
\titlespacing{\paragraph}{0pt}{3pt}{-1pt}

\newcolumntype{Y}{>{\centering\arraybackslash}X}    

\usepackage{tabu}

\begin{document}

\date{}                                         

\title{\vspace{-8mm}\textbf{\large
Flow-Acoustics: Theory and Benchmarking }}

\author{
Stefan Schoder  \\
\emph{\small Aeroakustik und Vibroakustik (AVG),
} \\
\emph{\small Institute of Fundamentals and Theory in Electrical Engineering (IGTE), 
} \\
\emph{\small TU Graz Inffeldgasse 16c, A-8010 Graz, Austria, Email: stefan.schoder@tugraz.at
} 
} \maketitle
\thispagestyle{empty}           

\section*{Introduction}

The urgent need for transitioning to green energy solutions, particularly in the context of house heating and urban redensification, has brought the issue of fan noise aeroacoustics investigations to the forefront. As societies worldwide strive to mitigate climate change and reduce carbon emissions, adopting sustainable heating technologies such as air heat pumps has gained significant traction. In Germany, renowned for its commitment to environmental sustainability, the "TA Lärm" regulations, derived from the "Bundes-Immissionsschutzgesetz," impose stringent limits on noise levels both inside and outside buildings across various applications \cite{TAL}. 
\begin{table}[ht]
\caption{"TA Lärm" limits in residential areas \cite{TAL}.}
\label{tab:TA-Lärm}
\begin{tabularx}{\linewidth}{@{}lll@{}}
    \toprule
    \multicolumn{3}{l}{\textbf{Immission guide values TA-Lärm}} \\
    \midrule
    \multicolumn{2}{l}{\textbf{Industrial Areas}} &  $70\,$dB(A) \\
    \multicolumn{3}{l}{\textbf{Core Areas, village areas and mixed areas}} \\
     & day & $60\,$dB(A) \\
     & night & $45\,$dB(A) \\
    \multicolumn{3}{l}{\textbf{Residential Areas}} \\
     & day & $50\,$dB(A) \\
     & night & $35\,$dB(A) \\
    \bottomrule
\end{tabularx}
\end{table}
These regulations, outlined in Tab.~\ref{tab:TA-Lärm}, delineate permissible noise levels during daytime (6 AM to 10 PM) and nighttime (10 PM to 6 AM), with particular emphasis on protecting residential areas with low noise limits. Moreover, the noise limits prescribed for indoor environments are even more stringent. Given the necessity of maintaining acoustic comfort and quality of life, compliance with these regulations necessitates meticulous attention to noise generation sources, especially those associated with heating and ventilation systems. Consequently, understanding and mitigating fan noise through aeroacoustic investigations is essential to ensure the successful adoption and integration of green energy solutions in residential and urban settings. In the following, an experimental benchmark for a low-pressure rise axial fan (FAN-01) is presented, and several prediction methods of the sound pressure and sound power are evaluated.

\section*{Experimental Reference: FAN-01}

The European Acoustics Association (EAA)-benchmark platform \cite{Hornikx2015,EAABenchmark} serves as a valuable resource, offering accessible benchmark data for validation purposes, managed by the Technical Committee of Computational Acoustics. Researchers are encouraged to utilize this platform to validate their calculations, report their findings, and contribute new benchmark data. The focus of this manuscript deals with the benchmark case FAN-01 involving a low-pressure axial fan, essential for sound design and optimization through robust aeroacoustic predictions using numerical methods. This benchmark data is akin to the Francis 99 concept \cite{Francis99,Turbine} and provides a download link and comprehensive data summary (Tab.~\ref{tab:summary}), with detailed experimental setup and measurement positions outlined for reference \cite{Zenger2016,schoder2022Fan}.

\begin{table*}[t]
\caption{Summary of the benchmark and the available data of the EAA low-pressure axial fan benchmark FAN-01.\protect\footnotemark}
\label{tab:summary}
\centering
\begin{tabularx}{\textwidth}{@{}lp{14cm}@{}}
\toprule
 Hosted by &  EAA \\
 Short Name & Dataset FAN-01\\
 Full Name & Benchmark-Dataset FAN-01: Low pressure Axial Fan in a short Duct\\
 Data: DOI & \url{https://doi.org/10.5281/zenodo.8418340}  \\ 
 Data-Paper: DOI & \url{https://doi.org/10.48550/arXiv.2211.12014}\\
 Citation &  S. Schoder, and Czwielong, F. (2022). Dataset FAN-01: Revisiting the EAA Benchmark for a low-pressure axial fan. arXiv preprint arXiv:2211.12014.\\
 Initial &  F. Zenger, C. Junger, M. Kaltenbacher, and
S. Becker. A benchmark case for aerodynamics
and aeroacoustics of a low pressure axial fan. In
SAE T.P. 2016-01-1805, 2016. \\
 URL &  \url{https://zenodo.org/records/8418340}\\ 
 Workshops & 1$^\text{st}$ at DAGA 2022 in Stuttgart
 \\
 \midrule
 \multicolumn{2}{l}{\textbf{Fan}}  \\
 Design parameters &  See Tab.~\ref{tab:fan_characteristics}  \\
 Geometry &  The rotor geometry is available as IGS-CAD model or Parasolid file.  \\
 Operating condition &  See Tab.~\ref{tab:fan_characteristics}  \\
 \midrule
 \multicolumn{2}{l}{\textbf{Laser Doppler Anemometry (LDA)}}  \\
 LDA system & 2-component LDA probe, type 2D FiberFlow
(Dantec Dynamics)
BSA P80 burst spectrum analyzer (Dantec Dy-
namics)
BSA Flow software v5.20 (Dantec Dynamics)\\
Measurement time & 12 min or $2.5\times10^6$ samples per position \\
\midrule
\multicolumn{2}{l}{\textbf{Wall pressure fluctuations}}  \\
Type & Differential pressure transducers XCS-093-1psi D
(Kulite Semiconductor Products) \\
Measurement time & 30 s with a sampling frequency of 48 kHz \\ 
\midrule
 \multicolumn{2}{l}{\textbf{Microphones}}  \\
 Type & 1/2 inch free-field microphones 4189-L-001 (Brüel
\& Kj\ae r) \\
Measurement time & 30 s with a sampling frequency of 48 kHz \\
\midrule
 \multicolumn{2}{l}{\textbf{Microphones (microphone array)}}  \\
Type & 1/4 inch array microphones 40PH-Sx (G.R.A.S.
Sound \& Vibration) \\
Measurement time & 30 s with a sampling frequency of 48 kHz \\
 \bottomrule
\end{tabularx}
\end{table*}

\begin{figure}[hbt]
    \centering
    \includegraphics[width=\linewidth]{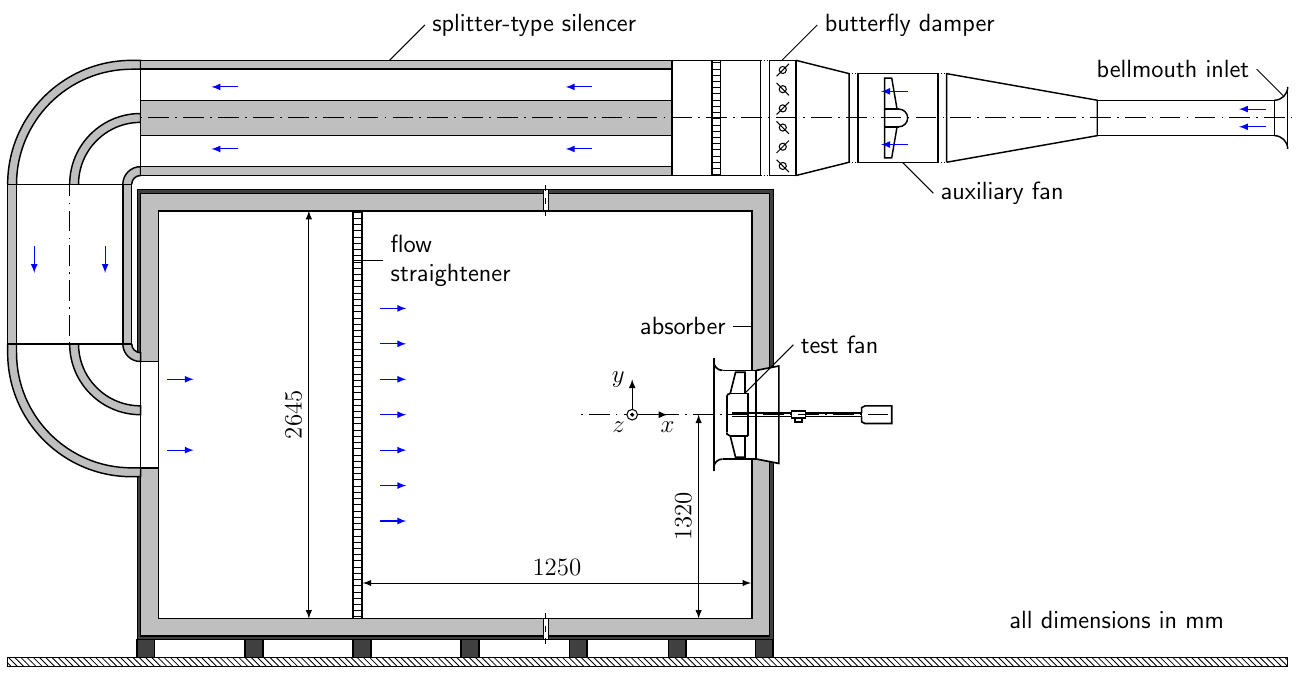}
    \caption{Standardized inlet test chamber.}
    \label{fig:schema}
\end{figure}

These data encompass aerodynamic performance metrics such as volume flow rate, pressure rise, and efficiency. Additionally, the dataset includes fluid mechanical parameters such as velocity in three spatial directions, turbulent kinetic energy acquired on both the fan suction and pressure sides, and wall pressure fluctuations within the duct (refer to Fig.~\ref{fig:schema}). Readers are directed to \cite{Zenger2016} for further details regarding the experimental configuration.
The fan, designed using blade element theory for low solidity fans, represents a typical model for industrial (HVAC) and automotive (thermal management cooling fan) applications, as outlined in Tab.~\ref{tab:fan_characteristics}. It features zero-blade skew and NACA 4510 profiles. The circumferential velocity at the blade tip corresponds to a Mach number of $0.113$, indicating an incompressible flow regime. Based on the chord length, the Reynolds number remains relatively constant across the blade span, signifying turbulent flow. Original geometry data is accessible online (refer to Tab.~\ref{tab:summary}). All measurements were conducted in a standardized anechoic inlet test chamber, adhering to ISO 5801 standards (see Fig.~\ref{fig:schema}).

\begin{table}[htb]
\caption{Fan design parameters (according to \cite{Zenger2016})}
\label{tab:fan_characteristics}
    \begin{tabularx}{\linewidth}{@{}ll@{}}
        \toprule
        Fan diameter & $495\,$mm \\
        Hub diameter & $248\,$mm \\
        Tip clearance & $2.5\,$mm \\
        Blades & 9 \\
        Rotational speed & $1486\,$min$^{-1}$ \\
        Chord length hub & $103\,$mm \\
        Chord length tip & $58\,$mm \\
        Reynolds number hub & $1.25\cdot 10^5$ \\
        Reynolds number tip & $1.50\cdot 10^5$ \\
        \bottomrule
    \end{tabularx}
\end{table}

\subsection*{Aerodynamic Performance Data}

Under operating conditions, the volumetric flow rate $\dot{V}$ was regulated using butterfly dampers and an auxiliary fan in the inlet section. While the measured pressure rise of the fan at the design point amounted to $\Delta p = 126.5\,$Pa, the design pressure difference was not entirely achieved due to unaccounted losses such as tip flow. The efficiency is $\eta= 53\,\%$ at the design point. Detailed information regarding the measurement setup can be found in \cite{Zenger2016}.

\subsection*{Geometry}
The CAD geometry is available online at the EAA benchmark platform \cite{EAABenchmark}.

\subsection*{Laser Doppler Anemometry}
\begin{figure}[hbt]
    \centering
    \includegraphics[width=0.9\linewidth]{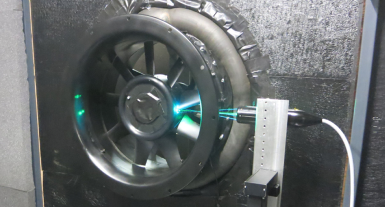}
    \caption{Picture of the LDA measurement series.}
    \label{fig:lda}
\end{figure}

Laser Doppler Anemometry (LDA) measurements (refer to Fig.~\ref{fig:lda}) provide flow velocity data, including meridional, radial, and circumferential components on both the suction and pressure sides. The available data include ensemble-averaged and time-averaged signals, which can be utilized to assess the turbulent kinetic energy on both sides and validate the turbulent structures within the inflow.

\footnotetext{Note that this repository was moved by the EAA from \url{https://www.tuwien.at/en/mwbw/mec/e325-03-research-unit-of-measurement-and-actuator-technology/eaa-benchmarks/benchmarks/acoustics-involving-heterogeneous-and-moving-fluids} to Zenodo during the recreation of the repository, when adding the dataset for FSAI \cite{FSAI,FSAI1}.}

\subsection*{Wall pressure measurements}
\begin{figure}[htbp]
    \centering
    \includegraphics[width=\linewidth]{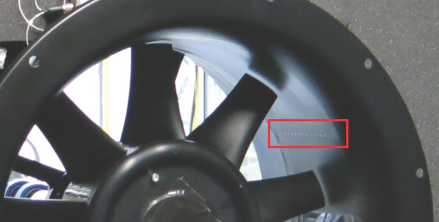}
    \caption{Picture of the wall pressure sensor distribution inside the duct.}
    \label{fig:Pres}
\end{figure}

The wall pressure fluctuation measurements consist of data from 15 transducers installed 15 mm after the end of the nozzle, with a spacing of 10 mm (see Fig.~\ref{fig:Pres}). The pressure probe measurements can validate the effects of the tip gap flow. 

\subsection*{Microphone}
\begin{figure}[htbp]
    \centering
    \includegraphics[width=\linewidth]{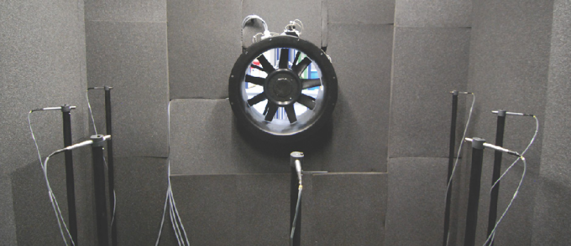}
    \caption{Picture of the microphone positions inside the anechoic measurement box.}
    \label{fig:Mic}
\end{figure}

The acoustic spectra at seven different microphone positions upstream of the fan are provided (see Fig.~\ref{fig:Mic}).

The available measurement results contain microphone signals with a 48\,kHz sampling frequency and measurement length of $T=30$\,s. Inside the test chamber, seven ($N=7$) microphones were installed horizontally in a half-circle with a radius of 1\,m in front of the nozzle
of the duct (see Fig.~\ref{fig:Mic}).
\begin{figure}[htbp]
    \centering
    \includegraphics[width=\linewidth]{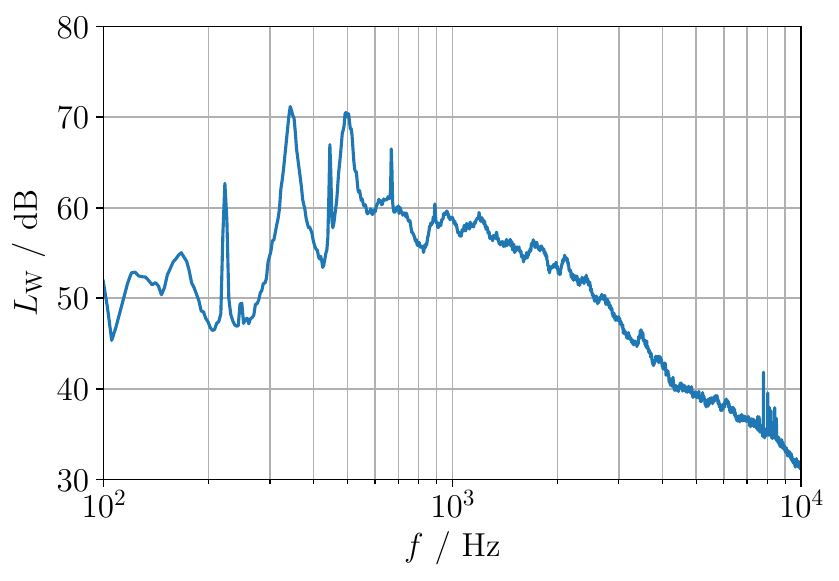}
    \caption{Measured sound power level of the fan.}
    \label{fig:soundpower}
\end{figure}
The sound power level was computed by
\begin{equation}
    L_{\rm W} = \bar{L}_{\rm P} + 10 \log\!\left(\frac{S_1}{S_0}\right) ~
    \text{dB}\,,
    \label{eq:soundpowerlevel}
\end{equation}
with the time-averaged sound pressure level $\bar{L}_{\rm P}$, the hull of
the measurement area $S_1=6.28\,$m$^2$ and $S_0=1\,$m$^2$. The time-averaged sound power level for all microphones is computed as
\begin{equation}
    \bar{L}_{\rm P} = 10 \log\!\left(\frac{1}{N}\sum_{n=1}^{N}
    \frac{1}{T}\int \frac{p_n^2}{p_0^2} \mathrm{d}t \right) ~ \text{dB}\,,
\end{equation}
with the reference pressure $p_0=20\,\mu$Pa.
For a frequency range of $100\,$Hz to $10\,$kHz, the measured sound power level was $L_{\rm W}=87.3\,$dB. The spectrum of the sound power level is shown in Fig.~\ref{fig:soundpower}.

\subsection*{Microphone Array}
\begin{figure}[hbt]
\centering
    \includegraphics[width=\linewidth]{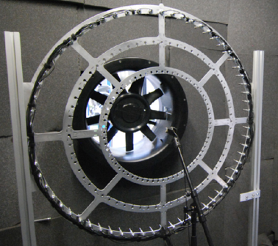}
\caption{Picture of the position of the microphone array relative to the axial fan.}
\label{fig:MicArray}
\end{figure}

Results for microphone array measurements (see Fig. \ref{fig:MicArray}) are provided in \cite{KroemerDiss}. 

\section*{Selected Prediction Methods}

\begin{figure}[hbt]
    \centering
    \includegraphics[width=\linewidth]{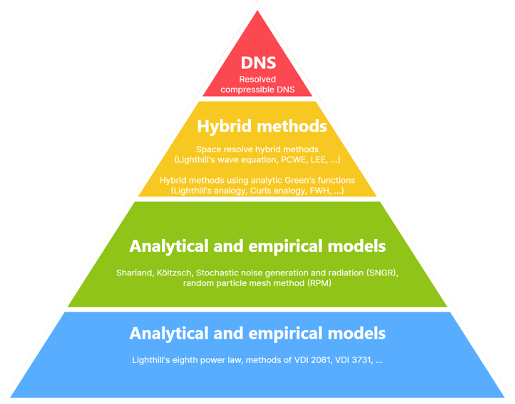}
    \caption{Hierarchy of aeroacoustic prediction methods ranging from \textit{class 1} (bottom) to \textit{class 4} (top).}
    \label{fig:Methods}
\end{figure}

With the trend towards greener heating, air-source heat pumps are increasingly used. One of the noise sources of air source heat pumps is the installed axial fan and, in connection with the increasing density of cities, the noise pollution of residents is increasing. The noise source is flow-acoustic and requires a flow-acoustic calculation. Systematically, a hierarchy of models with different physical accuracy and calculation efficiency can be derived (summarized in \cite{Junger2019} and illustrated in Fig.~\ref{fig:Methods}). 
In the simplest case, flow acoustics are described using analytical models in the form of scale models (\textit{class 1}), such as Lighthill's power law or the methods of VDI 2081 and VDI 3731 for technical noise emissions. \textit{Class 2} models incorporate empirical factors that allow a good prediction of the sound based on experience in specific situations. \textit{Class 3} models \cite{schoder} use a numerical decoupling strategy of the comparable expensive flow calculation (source) and from the acoustics and structure. Thus, this type of model describes a pure forward coupling. Finally, to numerically resolve the entire fluid-structure-acoustics interaction, the field equations are solved in a coupled manner (\textit{class 4}). Class 4 models are accurate and characterized by a very high computational effort. They represent the current research gold standard. The respective challenges and applications are explained using a typical axial fan.

\subsection*{Class 1: Empirical Methods}
Class 1 empirical methods try to estimate the sound power levels based on algebraic equations resulting from simple rules. As a consequence, only the basic fan properties can be accounted for, and no sound optimization of the source is possible.

\paragraph{VDI 2081}
The VDI 2081 \cite{VDI2081} standard's noise prediction is based on the noise law of Bommes \cite{Bommes}. The outlet duct sound power level $L_{\mathrm{W}}$ is estimated by
\begin{equation}
    L_{\mathrm{W}}=L_{\mathrm{WS}}+10 \log\!\left( \dot{V}\right)+5(\gamma-1) \log\!\left(\Delta p\right)
\end{equation}
which is a function of the volume flow rate $\dot{V}$, the total pressure increase $\Delta p$ between the inlet chamber and the ambient conditions, the specific sound power level $L_{\mathrm{WS}}$ and the Mach number exponent $\gamma$. The specific sound power level for axial fans can be estimated to be $L_{\mathrm{WS}} = 42$~dB (at the design point). The validity of these assumptions is questionable when the fan is not operated around the design point. For the given fan, $L_w= 85.6$~dB.

\paragraph{VDI 3731}
The noise prediction method outlined in part 2 of the VDI 3731 standard \cite{VDI3731}, proposed initially by Eck \cite{Eck}, assumes that the sound power $P$ is directly linked to the aerodynamic power loss and a power-law of the circumferential Mach number. It can be expressed like
\begin{equation}
P \propto \dot{V} \Delta p\left(\frac{1}{\eta_{\mathrm{i}}}-1\right)\left(\mathrm{Ma}_u\right)^m \, .
\end{equation}
Here, $\eta_{\mathrm{i}}$ denotes the inner efficiency (without losses from leakage and friction in bearings), $m$ represents the Mach number exponent, and $\mathrm{Ma}_u$ signifies the circumferential Mach number. The circumferential Mach number is given by
\begin{equation}
\mathrm{Ma}_u=\frac{u_a}{c_0}=\frac{\pi D n}{c_0} \, ,
\end{equation}
where $D$ stands for the outer fan diameter, $n$ denotes the rotational speed, and $c_0$ represents the speed of sound. This relationship reads logarithmically as
\begin{equation}
\begin{split}
    L_{\mathrm{W}}=L_{\mathrm{WS}} &+10 \log\!\left(\frac{\dot{V}}{\dot{V}_0} \frac{\Delta p}{\Delta p_0}\left(\frac{1}{\eta}-1\right)\right)\\
    &+10 m \log\!\left(\mathrm{Ma}_u\right) \, ,
\end{split}
\end{equation}
with reference values $\dot{V}_0 = 1$ m$^3$/s and $\Delta p_0 = 1$ Pa. Since measuring inner efficiency $\eta_{\mathrm{i}}$ is challenging, it is substituted with the total-to-static efficiency of the fan $\eta$. For axial fans, the specific sound power level can be approximated as $L_{\mathrm{WS}} = 96.6$~dB near the design point, and the exponent $m$ can be set to $3.16$. The total-to-static efficiency of the fan is calculated using
\begin{equation}
\eta=\frac{\dot{V} \Delta p}{2 \pi n M} \, ,
\end{equation}
where $M$ denotes the torque of the shaft. While the assumption of aerodynamic power loss as the source of acoustic power provides a conceptual link between the flow and acoustics. However, the direct correlation is debatable due to the significant disparity in magnitudes between aerodynamic losses and acoustic power. For the given fan, $L_w= 88.7$~dB.

\subsection*{Class 2: Effect-based Empirical Methods}
Class 2 empirical methods utilize algebraic equations to estimate sound power levels by summing up various effects derived from fundamental principles. The total noise is represented as a sum of individual noise sources, with required flow quantities typically obtained from averaged values like inflow velocity. These values can be derived from steady flow simulations or measurements and applied to different designs. These methods often assume simplified geometries, such as straight flat plates, reducing the generalization capabilities of these methods. Furthermore, these methods are limited to accounting for basic fan properties, prohibiting optimization of the sound source.

\paragraph{Sharland}
Sharland's method \cite{102} predicts the overall sound power by assuming blades to be flat and independently radiating without interference effects. Sharland categorizes three distinct noise sources: turbulent inflow (ti), pressure fluctuations in the turbulent boundary layer (tbl), and vortex shedding at the blade's trailing edge (vs). The sum of these contributions gives the total sound power
\begin{equation}
    P = P_{\mathrm{ti}} + P_{\mathrm{tbl}} + P_{\mathrm{vs}} \, .
\end{equation}
Each noise contribution is approximated by an integral from the inner radius $r_i$ to the outer radius $r_o$ of the blade
\begin{equation}
    \begin{aligned}
    P_{\mathrm{ti}} & \approx z \frac{1}{48 \pi} \frac{\rho}{c_0^3} \int_{r_i}^{r_o} l \Phi^2 w_{\infty}^6 \mathrm{Tu}^2 \mathrm{d}r \\
    P_{\mathrm{tbl}} & \approx 10^{-7} z  \frac{\rho}{c_0^3} \int_{r_i}^{r_o} l w_{\infty}^6 \mathrm{d}r \\
    P_{\mathrm{vs}} & \approx z \frac{1}{120 \pi} \frac{\rho}{c_0^3} \int_{r_i}^{r_o} l w_{\infty}^6 \operatorname{Re}^{-0.4} \mathrm{d}r \, .
    \end{aligned}
\end{equation}
The noise sources are influenced by the blade's chord length $l$ and the relative velocity $w_\infty$. The turbulent inlet noise contribution depends on parameters such as the turbulent intensity $\mathrm{Tu}$, fluid mass density $\rho$, bulk Reynolds number $\mathrm{Re}$, and the gradient of the lift coefficient $\Phi \approx 0.9\pi$ (an approximation by Sharland). The turbulent boundary layer noise contribution solely depends on the blade size. In contrast, the vortex shedding noise contribution depends on the Reynolds number $\mathrm{Re}$. However, this assumption regarding vortex shedding may be overly generalized, as it overlooks the influence of individual blade geometry. For blades with sharp trailing edges, this contribution may be small. Each contribution's sound power is multiplied by the number of blades $z$. This method predicts $L_w= 86.4$~dB for the given fan. 

\paragraph{Költzsch}
Költzsch's method \cite{66} predicts the power spectral density (PSD), offering spectral insights into the emitted sound. It identifies two primary sound sources: turbulent inflow (ti) and turbulent boundary layer (tbl), leading to a combined spectral density
\begin{equation}
    S = S_{\mathrm{ti}}(f) + S_{\mathrm{tbl}}(f) \, .
\end{equation}
This spectral density depends on frequency $f$. The PSD of turbulent inflow is approximated using the inflow's spectral energy density $S_w$, scaled by blade dimensions $l$, $b$, and the number of blades $z$
\begin{equation}
    S_{\mathrm{ti}}(f) = z \frac{0.81 \pi}{48} \frac{\rho}{c_0^3} w_{\infty}^4 S_w(f) l b \, . 
\end{equation}
In the same way as the method of Sharland, the turbulent inflow contribution uses an empirical constant and the proportionality to $\frac{\rho}{c_0^3}$. The inflow's spectral energy density, influenced by turbulent length scales, can be derived from measurements or Reynolds-averaged Navier-Stokes (RANS) simulations \cite{Junger2016}.
\begin{figure}[hbt!]
    \centering
    \includegraphics[width=\linewidth]{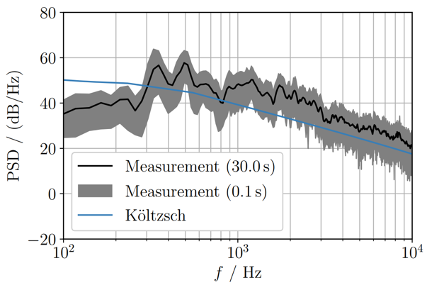}
    \caption{Results of the power spectral density of the Költzsch model. Furthermore, for comparison, measurements over 30~s are depicted in black, and measurements over 0.1~s are depicted in grey.}
    \label{fig:Koel}
\end{figure}
The power spectral density (PSD) of the turbulent boundary layer of a ducted fan is calculated as
\begin{equation}
    S_{\mathrm{tbl}}(f)=z \frac{\pi}{4} \frac{f}{\rho c_0^2 r_o\left(1-\nu^2\right)^2} S_{\mathrm{bl}}(f) \psi \, .
\end{equation}
Here, $S_{\mathrm{bl}}(f)$ represents the PSD of the lift forces on the blade, $\psi$ denotes a radiation function (approximated as 1 for low Mach number flows), and $\nu$ is the ratio of the outer diameter to the hub. The total noise contribution is obtained by multiplying it by the number of blades. The approximation of $S_{\mathrm{bl}}$ is provided in \cite{Junger2016,66}. Through these equations, Költzsch's method offers a spectral noise prediction, assuming turbulent inlet and turbulent boundary layer as sound sources (results see Fig. \ref{fig:Koel}).

\subsection*{Class 3: Hybrid Methods}
\begin{figure*}[htb!]
    \centering
    \includegraphics[width=1.0\textwidth]{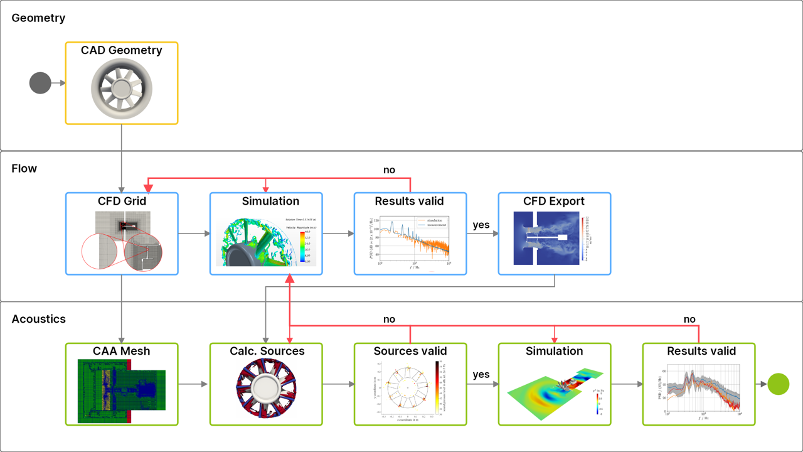}
	\caption{Validation workflow to perform hybrid aeroacoustic simulations \cite{hybrid,Flow} based on a step-by-step validation and a full integration of the benchmark data to validate the whole workflow using openCFS \cite{CFS,CFSDAT}.}
	\label{fig:Valid}
\end{figure*}
\begin{figure}[hbt!]
    \centering
    \includegraphics[width=\linewidth]{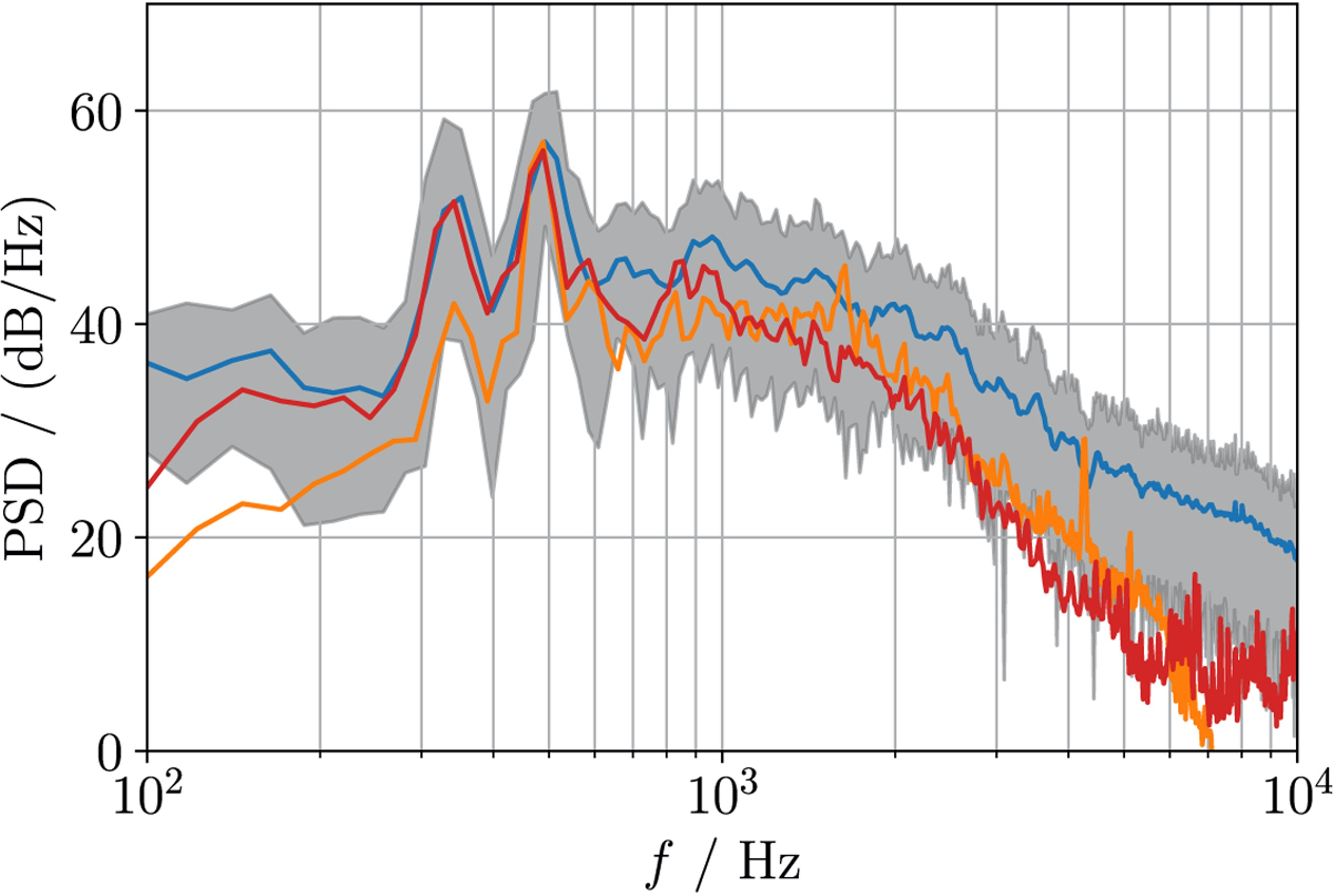}
	\caption{Results of the aeroacoustic simulations at microphone position 2, with the PCWE result in orange and the FWH result in red. Furthermore, for comparison, measurements over 30 s are depicted in blue, and measurements over 0.1 s are depicted in grey.}
	\label{fig:sim}
\end{figure}

Class 3 methods, like Lighthill's equation or the Perturbed Convective Wave Equation (PCWE) are derived from fundamental principles such as the perturbation ansatz. In doing so, the acoustic simulation is decoupled from the flow simulation. The detailed experimental and numerical findings of the EAA benchmark are compared in \cite{Fan01,Junger2019}, where the experimental results validate the entire aeroacoustic simulation process. The successful computation of flow and acoustic components follows a systematic procedure including validation steps. The flow grid is initially established based on a validated geometry, ensuring the mesh's accuracy for subsequent computational fluid dynamics (CFD) simulations. This process involves verifying turbulence modeling and conducting a flow convergence study using the grid convergence index \cite{Fan7}. Flow results' validation encompasses global and unsteady local quantities, particularly those related to the aeroacoustic source term in the wave equation, such as incompressible flow pressure for the perturbed convective wave equation (PCWE). After confirming the flow's validity, source calculation and acoustic simulation are performed. The computed sources are transformed while conserving energy to the acoustic mesh \cite{schoder}, with further validation conducted using microphone array measurements. Once the sources and acoustic mesh are validated, the acoustic computation proceeds, culminating in the final validation of the overall workflow through comparison with experimental results. Further details on the validation process for each step are outlined in \cite{Junger2019}. Figure \ref{fig:sim} presents the final validation plot, illustrating the comparison between experimental results, the PCWE ($L_w= 86.7$~dB), and the Ffowcs Williams and Hawkings (FWH) analogy ($L_w= 85.2$~dB).

\subsection*{Class 4: Simulation of the Compressible Flow \cite{Fan9}}
Finally, class 4 methods directly solve the compressilbe fully resolved flow equations using numerical techniques. In \cite{Fan9}, the LBM/VLES approach proved highly effective in forecasting the aerodynamic and aeroacoustic performance of the low-pressure axial fan. The numerical setup was found by a grid independence study, affirming the convergence quality of both aerodynamic and aeroacoustic parameters. Despite the considerable computational expense of the finest grid, amounting to about 1300 CPU hours per fan revolution, the approach exhibited satisfactory outcomes even with a coarser mesh, demanding only 175 CPU hours/rev. On a 308-core cluster, the wall clock time for fine and coarse simulations was approximately 84 and 11 hours, respectively, making this approach exceptionally competitive regarding industrial turnaround times.

\begin{figure}
    \centering
    \includegraphics[width=\linewidth]{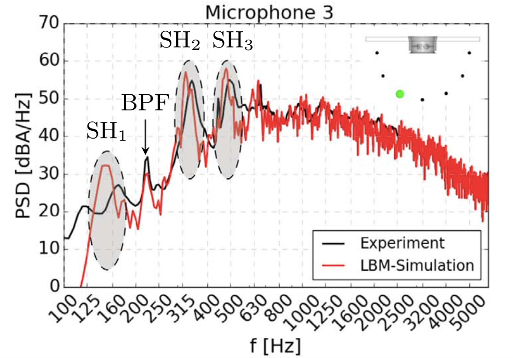}
	\caption{Results of the aeroacoustic simulations at the respective microphone position 3 in red. Furthermore, for comparison, measurements over 30 s are depicted in black, and measurements over 0.1 s are depicted in grey.}
	\label{fig:DNS}
\end{figure}

The computed flow metrics (pressure rise, fan efficiency), the axial velocity component, and predictions of wall-pressure fluctuations aligned well with experimental findings. Also, the far-field noise levels matched both broadband and tonal components (see Fig. \ref{fig:DNS}, $L_w= 86.7$~dB). A thorough analysis delved into the noise generation mechanisms within the tip clearance, revealing intricate interactions between turbulent structures, large coherent vortices, and the fan blades. These interactions led to the emission of broadband noise due to the continuous impingement of turbulent structures on the blades, as well as the generation of sub-harmonic humps resulting from the interaction of coherent tip vortices with the following blades. A similar investigation was presented in \cite{Fan4}.

\section*{Summary}
Overall, a selection of sound prediction models was presented. In its simplest form, flow acoustics are typically addressed using analytical models falling under class 1, such as Lighthill's power law or methods outlined in VDI 2081 and VDI 3731 for technical noise emissions. Class 2 models enhance these by incorporating empirical factors, drawing from specific experiential knowledge to provide sound predictions tailored to particular scenarios. Class 3 models, as described by Schoder et al. \cite{schoder}, adopt a numerical decoupling approach, separating the computationally intensive flow calculation (source) from the acoustic and structural aspects. This strategy entails a pure forward coupling method. Class 4 models tackle the comprehensive fluid-structure-acoustics interaction by concurrently solving the field equations, representing the pinnacle of accuracy, albeit with significantly increased computational demands. These models serve as the current benchmark in research and are characterized by their ability to resolve complex interactions thoroughly. Only the class 3 and 4 models can be reliably used for optimization of axial fan noise emission (as first investigations show in connection with machine learning techniques \cite{Fan13}). Furthermore, the transient operating point effects of ducted fans may be worth investigating in the future \cite{Fan10}.

\section*{Acknowledgement}
Thanks to C.~Junger for his excellent collaboration and support in his field of expertise, and to F.~Kraxberger for proofreading.

\bibliographystyle{plain}

\end{document}